# D-MG Tradeoff and Optimal Codes for a Class of AF and DF Cooperative Communication Protocols

Petros Elia, K. Vinodh, M. Anand and P. Vijay Kumar, *Fellow, IEEE*

***Abstract*— We consider cooperative relay communication in a fading channel environment under the Orthogonal Amplify and Forward (OAF) and Orthogonal and Non-Orthogonal Selection Decode and Forward (OSDF and NSDF) protocols.**

**For all these protocols, we compute the Diversity-Multiplexing Gain Tradeoff (DMT). We construct DMT optimal codes for the protocols which are sphere decodable and, in certain cases, incur minimum possible delay.**

**Our results establish that the DMT of the OAF protocol is identical to the DMT of the Non-Orthogonal Amplify and Forward (NAF) protocol.**

**Two variants of the NSDF protocol are considered: fixed-NSDF and variable-NSDF protocol. In the variable-NSDF protocol, the fraction of time duration for which the source alone transmits is allowed to vary with the rate of communication. Among the class of static amplify-and-forward and decode-and-forward protocols, the variable-NSDF protocol is shown to have the best known DMT for any number of relays apart from the two-relay case. When there are two relays, the variable-NSDF protocol is shown to improve on the DMT of the best previously-known protocol for higher values of the multiplexing gain. Our results also establish that the fixed-NSDF protocol has a better DMT than the NAF protocol for any number of relays.**

**Finally, we present a DMT optimal code construction for the NAF protocol.**

## I. Introduction

Cooperative relay communication is a promising means of wireless communication in which cooperation is used to create a virtual transmit array between the source and the destination, thereby providing the much-needed diversity to combat the fading channel.

Consider a communication system as shown in Fig. 1, in which there are $n+1$ nodes that cooperate in the communication between source node $S$ and destination node $D$. The remaining $(n-1)$ nodes thus act as relays.

Petros Elia and P. Vijay Kumar are with the Department of EE-Systems, University of Southern California, Los Angeles, CA 90089 USA (email: {elia,vijayk}@usc.edu).

K. Vinodh and M. Anand are with the Department of Electrical Communication Engineering, Indian Institute of Science, Bangalore, 560 012 India (email: {kvinodh,manand}@ece.iisc.ernet.in).

The results in this paper were presented in part at the 43rd Allerton Conference on Communications, Control and Computing, Sept. 2005, and in part at MILCOM 2006.

This work was carried out while P. Vijay Kumar was on leave of absence at the Indian Institute of Science, Bangalore.

This research is supported in part by NSF-ITR CCR-0326628 and in part by the DRDO-IISc Program on Advanced Research in Mathematical Engineering.

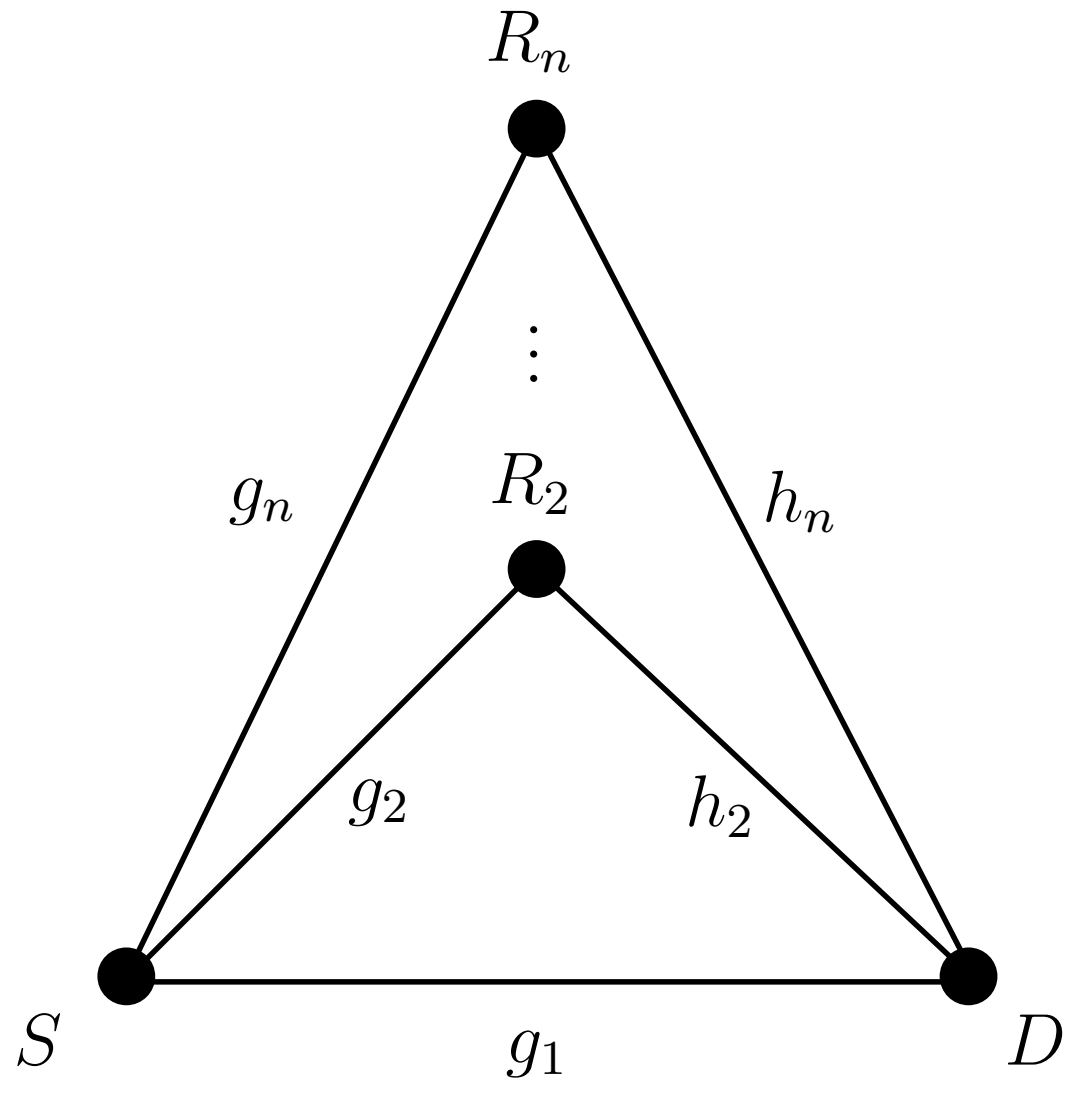

Fig. 1. Cooperative Relaying in networks.

### A. Assumptions

We follow the literature in making the assumptions listed below concerning the channel. Our description is in terms of the equivalent complex-baseband, discrete-time channel.

- All nodes have a single transmit and single receive antenna and are assumed to transmit synchronously.
- The $T$ channel uses over which communication takes place is short enough to invoke the quasi-static assumption, i.e., the channel fading coefficients $\{g_i, h_j\}$ are fixed for the duration of the communication, but vary randomly from one block of $T$ channel uses to the next. Here, $g_1$ denotes the fading coefficient for the channel between $S$ and $D$, $g_i$ for the channel between $S$ and the $i$th relay node $R_i$, and $h_i$ for the channel between $R_i$ and $D$, where $2 \leq i \leq n$.
- All channels are assumed to be Rayleigh fading and all fade coefficients are i.i.d., circularly symmetric complex gaussian $\mathbb{CN}(0,1)$ random variables.
- We assume half-duplex operation at each node, i.e., at any given instant a node can either transmit or receive, but not do both.
- The noise vector at the receivers is assumed to be comprised of i.i.d., circularly symmetric complex gaussian $\mathbb{CN}(0,\sigma^2)$ random variables.
- The destination must know all the fading coefficients in the case of the AF protocols and it needs to know only the

source-destination and relay-destination channel fading coefficients in the case of the DF protocols. We assume that a relay knows only the corresponding source-relay channel fading coefficient, i.e., relay $R_j$ will know only $g_j$ in the case of the DF protocols.

### B. Channel Model

We propose several cooperative communication protocols in this paper. All the protocols considered here involve two-phase communication. In the first phase, lasting for $p$ channel uses, the source $S$ broadcasts to the relays and the destination. In the second phase, lasting for $q$ channel uses, the relays communicate with the destination. A protocol is said to be non-orthogonal or orthogonal depending on whether the source continues to transmit (to the destination) in the second phase or otherwise. It is said to be a decode-and-forward (DF) or amplify-and-forward (AF) protocol depending upon whether the relays are required to decode the received message or not.

As we shall see, one can associate with every protocol, a channel model of the form

$$\underline{y} \ = \ H\underline{x} + \underline{w} \ , \tag{1}$$

where $\underline{y}$ corresponds to the received signal, $\underline{w}$ is the noise vector, $H$ is the virtual channel matrix and $\underline{x}$ is the vector transmitted by the source in the case of the AF protocols, and is the compound vector formed by concatenating the transmissions of the source and the participating relays in case of the DF protocols.

### C. Diversity-Multiplexing Gain Tradeoff

The average signal-to-noise ratio of a channel will be denoted by $\rho$. In the channel model in (1), the communication has happened over $m = p + q$ channel uses.

The probability of outage for the channel in (1) is defined as

$$P_{\text{out}}(R) = \inf_{\Sigma_x \ \geq \ 0, \ \text{Tr}(\Sigma_x) \ \leq \ p\mathcal{E}} \Pr(I(\underline{x};\underline{y}|H) \ \leq \ mR),$$

where $R$ is the average rate of communication. We have imposed an average energy constraint on the transmitted signal $\underline{x}$ by upper bounding the trace of its covariance matrix $\Sigma_x$. As will be shown later, by normalizing the noise vector, we can regard $\mathcal{E}$ as $\rho$.

In this paper, we use the symbol $\doteq$ to denote exponential equality, i.e., the expression

$$\lim_{\rho\to\infty} \frac{\log f(\rho)}{\log \rho} = b$$

is denoted by $f(\rho) \doteq \rho^b$ and $\dot{\geq}$, $\dot{\leq}$ are similarly defined.

We define $d_{\text{out}}(r)$ to be the negative $\rho$ exponent of $P_{\text{out}}(r \log \rho)$, where $R = r \log \rho$, i.e.,

$$P_{\text{out}}(r) \doteq \rho^{-d_{\text{out}}(r)}.$$

A MIMO coding scheme $\{C(\rho)\}$ for the channel model in (1) is said to achieve a spatial multiplexing gain $r$ if

$$\lim_{\rho\to\infty} \frac{R(\rho)}{\log \rho} \ = \ r$$

where $R(\rho)$ is the rate of the code $C(\rho)$. The coding scheme $\{C(\rho)\}$ is said to achieve a diversity gain $d(r)$ if

$$\lim_{\rho\to\infty} \frac{\log P_e(\rho)}{\log \rho} \ = \ -d(r) \ ,$$

where $P_e(\rho)$ is the average error probability of the code $C(\rho)$ under maximum likelihood decoding. Thus

$$P_e(\rho) \doteq \rho^{-d(r)}.$$

It can be shown from an application of Fano's inequality (see [2]) that

$$d(r) \leq d_{\text{out}}(r).$$

The diversity-multiplexing gain tradeoff (DMT) introduced by Zheng and Tse in [2], provides a means of evaluating and comparing the various proposed protocols. In this paper, we will by abuse of terminology refer to the outage exponent $d_{\text{out}}(r)$ as the DMT of the corresponding channel even though strictly speaking it is the DMT only if one can identify a code whose diversity gain $d(r)$ is equal to $d_{\text{out}}(r)$. No loss of accuracy is entailed in this because, as we shall see, for every protocol discussed in this paper we are able to identify a correspondingly optimal coding scheme. For this reason, we often will write $d(r)$ in place of the more accurate, but more cumbersome notation $d_{\text{out}}(r)$.

For each of the protocols described in the paper, the relation between received and transmitted vectors can alternately be written in the form

$$Y = HX + W \ ,$$

where $H$ is a row matrix whose components represent the fading coefficients a MISO channel and where $X$ is a code matrix, constrained by the particular protocol, to have a prescribed format. From this, it follows that the probability of outage of the channel model in (1) is at least as large as that of the corresponding MISO channel. Since the MISO channel is known to have outage exponent given by

$$d(r) = n(1 - r) \ ,$$

when $(n - 1)$ is the total number of relays, the quantity $d(r)$ is referred to in the literature as the transmit diversity bound [4].

### D. Prior Work in Cooperative Communication Protocols

The concept of user cooperative diversity was introduced in [12], [13]. Cooperative diversity protocols were first discussed in [4], where the authors develop and analyze the Orthogonal Amplify and Forward (OAF) protocol and the Selection Decode and Forward (SDF) protocol for the case of a single relay. In [6], the SDF protocol is analyzed for an arbitrary number of relays, where the authors give upper and lower bounds on the DMT of the protocol. In these protocols, the relays and the source node participate for equal time instants and the maximum multiplexing gain $r$ that could be achieved was $0.5$.

In [5], Azarian *et al.* analyze the class of Non Orthogonal amplify and Forward (NAF) protocols, introduced earlier by Nabar *et al.* in [11]. In [5], the authors establish the improved DMT of the NAF protocol in comparison to the class of OAF protocols considered in [4]. The authors also introduce the Dynamic Decode and Forward (DDF) protocol wherein the time for which the relays listen to the source depends on the source-relay channel gain. They show that the DMT of the DDF protocol achieves the transmit diversity bound for $r \leq 0.5$, beyond which the DMT falls below the transmit diversity bound.

Jing and Hassibi [7] consider cooperative communication protocols where the relay nodes apply a linear transformation to the received signal. The authors consider the case when both the source and the relays transmit for an equal number of channel uses and the linear transformation applied by the relays are restricted to the class of unitary matrices.

Yang and Belfiore consider a class of protocols called Slotted Amplify And Forward (SAF) protocols in [10], and show that these improve upon the performance of the NAF protocol [5] for the case of two relays. The authors also provide an upper bound on the DMT of the SAF protocol with any number of slots, and show that this upper bound tends towards the transmit diversity bound as the number of slots increases. Under the assumption of relay isolation and relay ordering, the naive SAF scheme proposed in [10] is shown to achieve the SAF protocol upper bound.

Yuksel and Erkip in [18] have considered the DMT of the DF and compress-and-forward (CF) protocols. They show that the CF protocol achieves the transmit diversity bound for the case of a single relay. We note however, that in the CF protocol, the relays are assumed to know all the fading coefficients in the system.

Recent work directed towards the construction of explicit coding schemes which are DMT optimal for cooperative protocols can be found in [9] and [10].

In the present paper, neither relay ordering nor relay isolation is assumed.

### E. Results

The results presented in this paper pertain to four protocols.

*1) Orthogonal Amplify and Forward:* This protocol was introduced by Laneman *et al.* [4], who analyze this protocol for the case of a single relay. As the name of the protocol suggests, in this two-phase protocol, the source broadcasts for $p$ time slots, followed by a relaying phase, lasting for $q$ time slots, where the relays amplify and forward the signals received by them.

Our version of the protocol is slightly more general than that in [4] since we permit the relays to operate on the received signal using a linear transformation[1] and we allow the source and the relays to transmit for unequal time slots. For this more general version of the protocol, we are able to determine the best possible DMT as well as construct, in a simple way, DMT optimal codes that incur minimum delay. Our results establish that the DMT of the OAF protocol matches the best possible DMT of the NAF protocol [5].

[1] In this sense, a more appropriate term for this protocol we consider here might be "linearly transform and forward". However we shall use the more generic term OAF here.

*2) Selection Decode and Forward:* This class of protocols was introduced by Laneman and Wornell [6]. In this protocol, in the first phase occupying $p$ channel uses, the source broadcasts to the destination and the relays. In the second phase, all the relays which are not in outage decode the source message, separately encode and transmit to the destination over the course of $q$ channel uses. Once again, we consider the case when $p \neq q$.

We consider two versions of this protocol in the present paper:

- *Non-Orthogonal Selection Decode and Forward (NSDF)* In this protocol, the source continues to transmit during the second phase. At this point, it is convenient to distinguish between two variants of this protocol.
  - *Variable-NSDF* In this variant, in order to obtain the best DMT possible, we allow $p$ and $q$ to vary with the multiplexing gain $r$. Note that since, $p$ and $q$ are not a function of the channel fading coefficients, this protocol still falls within the category of static protocols.
    - Among the class of static DF and AF protocols, the variable-NSDF protocol has the best-known DMT for any number of relays, except for the case of two relays.
    - The DMT of the variable-NSDF protocol for the case of two relays is better than the tradeoff of the SAF protocol [10] for $r > 0.6$ (refer Fig. 3).
  - *Fixed-NSDF* Here $p$ and $q$ are fixed and independent of $r$. The DMT for this variant of the NSDF protocol is also presented here for every pair $p, q,\ p \geq q$.
    - For values of the ratio $\kappa = \frac{p}{q}$ in the range $1 < \kappa < \frac{n}{n-1}$, the fixed-NSDF protocol dominates the NAF protocol for $0 \leq r \leq \frac{p}{p+q}$, beyond which they both have the same DMT.
- *Orthogonal Selection Decode and Forward (OSDF)* In this protocol, the source remains silent in the second phase. As in the case of the NSDF protocol, there are two variants of this protocol, fixed and variable. Under the variable-OSDF protocol, we allow $p$ and $q$ to vary with $r$ in order to compute the best DMT. We state the DMT for both the fixed and variable-OSDF protocols.
  The DMT of the variable-OSDF protocol, for the case of two relays, improves on the tradeoff of the SAF protocol [10] for $r > \frac{5}{8}$. The DMT of the variable-NSDF protocol is, however, better than the DMT of the variable-OSDF protocol for all $r$ for any number of relays.

All the OAF, NSDF and OSDF protocols considered in this paper are static protocols. In Fig. 2 and Fig. 3, we have shown the optimal DMT of the OAF, NSDF and OSDF protocols for the case of one and two relays respectively. In the figures, we have also shown the DMT of the NAF and SAF protocols for the sake of comparison. Note that the NAF and SAF protocols have the same DMT for the case of a single relay.

As our final result, we present a code that achieves the DMT of the NAF protocol considered by Azarian et. al. [5]. This result, first presented in [17], is included here for the sake of completeness.

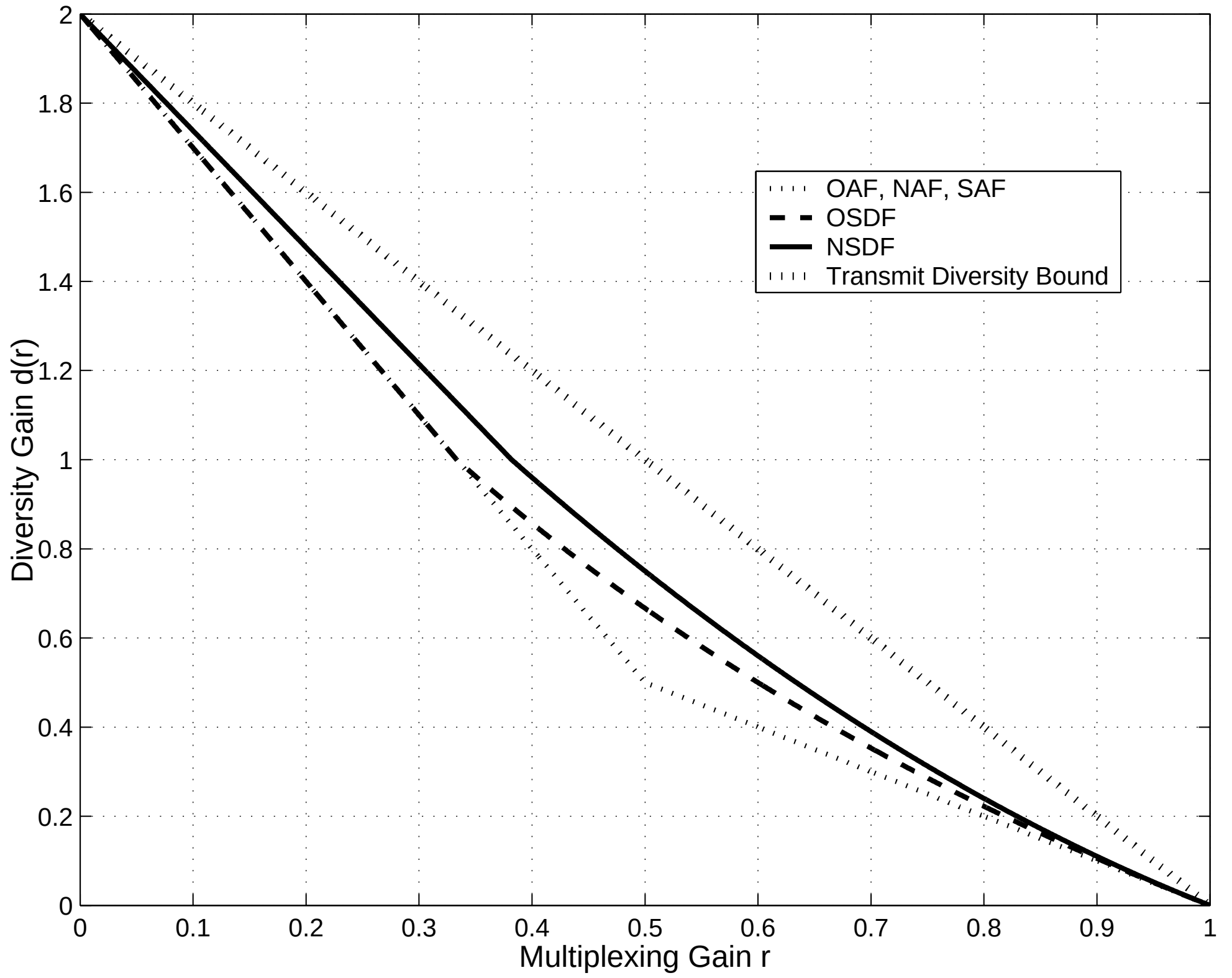

Fig. 2. *Optimal DMT for single relay cooperative communication protocols.*

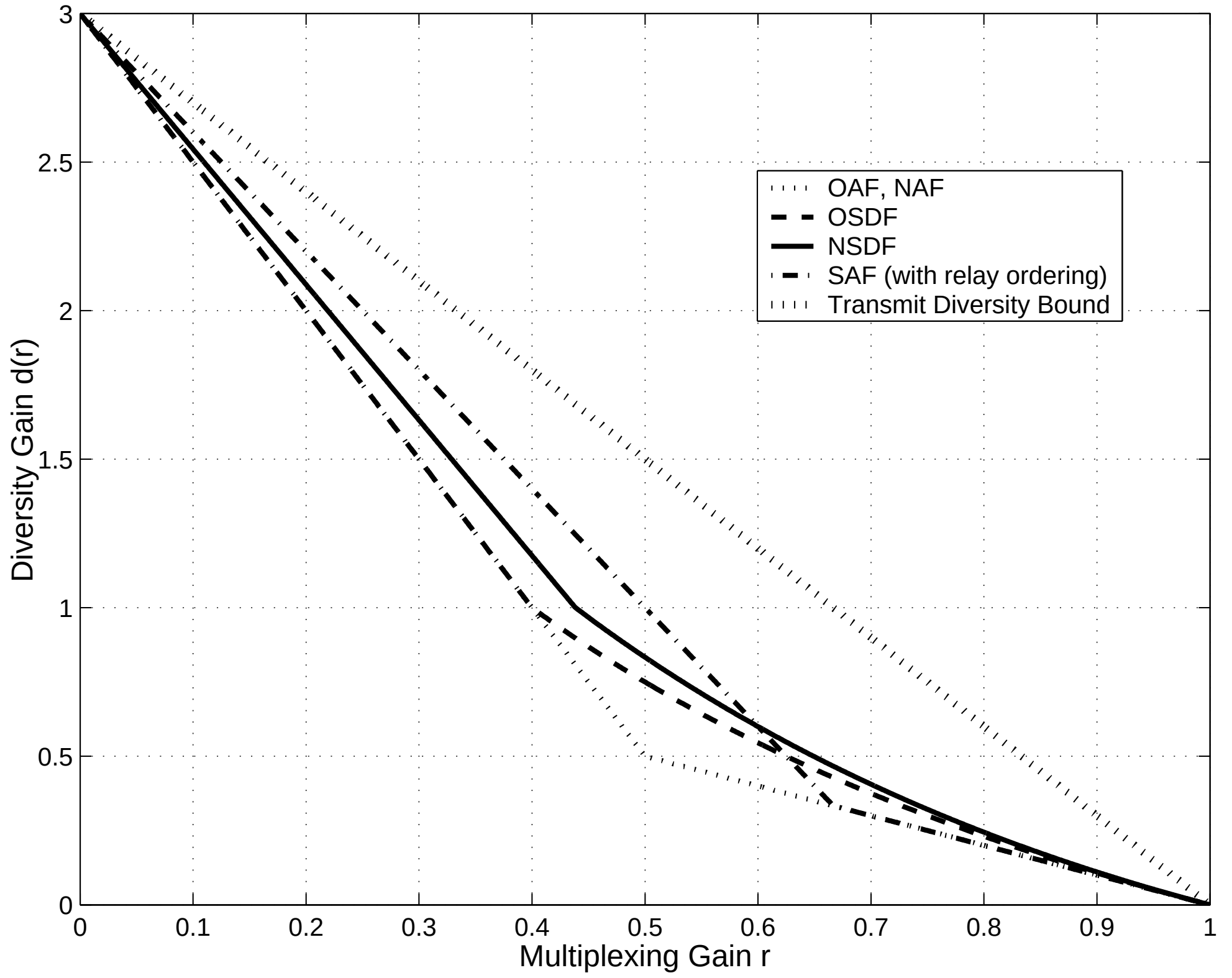

Fig. 3. *Optimal DMT for two relay cooperative communication protocols.*

### F. Organization of the Paper

In Section II, we discuss the class of OAF protocols and compute their DMT. Section III contains the description and DMT analysis of the NSDF protocols. We state the DMT of the OSDF protocols in Section IV. We also present constructions for DMT optimal codes, based on cyclic division algebras, for the OAF, NSDF and OSDF protocols in the corresponding sections. Finally, we provide a code construction which is DMT optimal for the NAF Protocol [5] in Section V.

*Notation:* The norm of a vector and the Frobenius norm of a matrix are denoted by $\|.\|$ and $\|.\|_F$ respectively. $|.|$ will denote the determinant of a matrix as well as the modulus of a complex scalar.

## II. Orthogonal Amplify and Forward Protocol

In this protocol, the source $S$ transmits a signal to the relays $\{R_j\}$ and to the destination $D$ for $p$ channel uses. Over the next $q$ channel uses, the relays transmit a linear transformation of the received signal, while the source remains silent.

As it turns out, for $\frac{p}{p+q} = \frac{n}{2n-1}$, we get the best DMT for all $r$ in the range $0 \leq r \leq 1/2$. For $r > 1/2$, we obtain the best DMT when there is no cooperation from the relays and the source continuously transmits to the destination.

Hence, we get the DMT of the OAF protocol by choosing $\frac{p}{p+q} = \frac{n}{2n-1}$ when $r \leq 1/2$ and $\frac{p}{p+q} = 1$ when $r > 1/2$.

### A. OAF Channel Model

Based on the above signalling protocol, we have the following model for the received signal.

$$\begin{aligned}
\underline{y}_1 &= g_1\underline{x} + \underline{w}_1 \\
\underline{r}_j &= g_j\underline{x} + \underline{v}_j \ , \ j = 2,3,\ldots,n \\
\underline{y}_2 &= \sum_{j=2}^{n} h_j A_j \underline{r}_j + \underline{w}_2 \\
&= \left[\sum_{j=2}^{n} g_j h_j A_j\right]\underline{x} + \sum_{j=2}^{n} h_j A_j \underline{v}_j + \underline{w}_2 \ ,
\end{aligned}$$

where $\underline{x}$ is the signal transmitted by the source, $[\underline{y}_1^t, \underline{y}_2^t]^t$ is the signal received by the destination and $\underline{r}_j$ is the signal received by the $j$th relay. $\{A_j\}$ are $(q \times p)$ complex matrices that represent the linear transformation at the relay nodes. The vectors $\{\underline{v}_j\}_{j=2}^n$ and $\{\underline{w}_1, \underline{w}_2\}$, represent additive noise seen at the relay nodes and the destination respectively. The above channels can be rewritten in matrix format as

$$\underline{y} = H\underline{x} + \underline{n}, \tag{2}$$

where

$$\begin{aligned}
\underline{n} &= \begin{bmatrix} \underline{w}_1 \\ \sum_{j=2}^n h_j A_j \underline{v}_j \ + \ \underline{w}_2 \end{bmatrix} \\
H &= \begin{bmatrix} g_1 I_p \\ \sum_{j=2}^n g_j h_j A_j \end{bmatrix} \\
\underline{y} &= \begin{bmatrix} \underline{y}_1 \\ \underline{y}_2 \end{bmatrix}.
\end{aligned}$$

The covariance matrices of the noise and signal vector are denoted by

$$\begin{aligned}
\Sigma_n &:= \mathbb{E}(\underline{nn}^\dagger) \\
&= \begin{bmatrix} \sigma_w^2 I_p & 0 \\ 0 & \sigma_w^2 I_q \ + \ \sum_{j=2}^n |h_j|^2 \sigma_{v_j}^2 A_j A_j^\dagger \end{bmatrix}
\end{aligned}$$

and

$$\Sigma_x := \mathbb{E}(\underline{xx}^\dagger),$$

where $\sigma_{v_j}^2, \sigma_w^2$ denote the variances of the corresponding noise vectors. We impose the energy constraint,

$$\mathrm{Tr}(\Sigma_x) \leq p\mathcal{E},$$

where $\mathcal{E}$ denotes the average energy available for transmission of a source symbol. The SNR $\rho$ is defined as

$$\rho := \frac{\mathcal{E}}{\sigma_w^2} \tag{3}$$

We assume the ratio of the noise variances $\{\sigma_{v_j}^2\}$ and $\sigma_w^2$ to be a constant independent of $\rho$. The average energy of the signal transmitted by the $j^{th}$ relay is given by,

$$\begin{aligned}
&\mathbb{E}\{\|g_j A_j \underline{x} + A_j \underline{v}_j\|^2\} \\
&\quad = |g_j|^2 \mathbb{E}\{\|A_j \underline{x}\|^2\} + \mathrm{Tr}(A_j \mathbb{E}\{\underline{v}_j \underline{v}_j^\dagger\} A_j^\dagger) \\
&\quad \leq |g_j|^2 p\mathcal{E} \mathrm{Tr}\{A_j A_j^\dagger\} + \sigma_w^2 \mathrm{Tr}\{A_j A_j^\dagger\} \\
&\quad \leq \alpha_j^2 (p\mathcal{E}|g_j|^2 + \sigma_w^2)
\end{aligned}$$

where $\alpha_j^2$ is the squared Frobenius norm of the relay matrices $\{A_j\}$, i.e.,

$$\|A_j\|_F^2 \ = \ \mathrm{Tr}(A_j A_j^\dagger) \ := \ \alpha_j^2. \tag{4}$$

We have used the inequality

$$\Sigma_x \ \leq p\mathcal{E} I_p$$

in coming up with the above bound.

We impose the constraint that the average energy transmitted by a relay satisfy

$$\mathbb{E}\{\|g_j A_j \underline{x} + A_j \underline{v}_j\|^2\} \leq \mathcal{E},$$

and this is achieved by requiring that

$$\alpha_j^2 \leq \frac{\mathcal{E}}{p\mathcal{E}|g_j|^2 + \sigma_w^2}. \tag{5}$$

As in [5], the SNR exponent of $\alpha_j^2$ can be made zero. In practice, we can choose $\alpha_j$ to be a suitable constant so that the probability of the event described by (5) is high.

### B. Upper bound on the DMT of the class of OAF protocols

We first state a useful lemma concerning non-negative definite matrices.

*Lemma 1:* Let $\{g_j, h_j, A_j\}$ be as defined above. Let

$$B = \sum_{j=2}^{n} g_j h_j A_j.$$

Then,

$$BB^\dagger \;\leq\; (n-1)\sum_{j=2}^{n}\gamma_j A_j A_j^\dagger$$

as non-negative definite matrices, where $\gamma_j = |g_j h_j|^2$.

*Proof:* Please see Appendix I. ■

Next, we establish the upper bound on the DMT of the class of OAF protocols.

*Theorem 2: (General OAF Upper Bound)* Consider the collection of OAF protocols described above (different protocols can be obtained by varying $p$, $q$ and $\{A_j\}$ for a given $n$). Then, regardless of the choice of the transformation matrices $\{A_j\}$, the DMT of any protocol satisfies the upper bounds given below.

If $\frac{p}{m} \geq \frac{n}{2n-1}$, where $m = p+q$,

$$d(r) \;\leq\; \begin{cases} n\left(1-\frac{(n-1)mr}{nq}\right), & 0 \leq r \leq \frac{q}{m} \\ \frac{p}{p-q}\left(1-\frac{mr}{p}\right), & \frac{q}{m} \leq r \leq \frac{1}{2} \\ (1-r), & \frac{1}{2} \leq r \leq 1 \end{cases}. \tag{6}$$

If $\frac{p}{m} \leq \frac{n}{2n-1}$, then

$$d(r) \;\leq\; \begin{cases} n(1-\frac{mr}{p}), & 0 \leq r \leq \frac{(n-1)}{n\frac{m}{p}-1} \\ (1-r), & \frac{(n-1)}{n\frac{m}{p}-1} < r \leq 1 \end{cases}. \tag{7}$$

In deriving these bounds for the protocols, cooperative relaying is avoided whenever it is advantageous to do so.

Also, the highest value of the upper bound on the DMT occurs for the choice $\frac{p}{m} = \frac{n}{2n-1}$. In this case, we get

$$d(r) \;\leq\; \begin{cases} n\left(1-\frac{(2n-1)r}{n}\right), & 0 \leq r \leq \frac{1}{2} \\ (1-r), & \frac{1}{2} < r \leq 1 \end{cases}. \tag{8}$$

*Proof:* As in [2], the input distribution can be assumed to be circularly symmetric gaussian without loss of generality. The maximum mutual information between $\underline{x}$ and $\underline{y}$, conditioned on the knowledge of $H$ at the receiver, is given by

$$\begin{aligned} \mathcal{I}_{\max} &= \max_{\Sigma_x \,\geq\, 0,\ \mathrm{Tr}(\Sigma_x) \,\leq\, p\mathcal{E}} I(\underline{x};\underline{y} \mid H) \\ &= \max_{\Sigma_x \,\geq\, 0,\ \mathrm{Tr}(\Sigma_x) \,\leq\, p\mathcal{E}} \log | I_m + H\Sigma_x H^\dagger \Sigma_n^{-1} | . \end{aligned}$$

Arguing as in [2], in the scale of interest we get

$$\mathcal{I}_{\max} \;\doteq\; \log | I_m + \mathcal{E}HH^\dagger\Sigma_n^{-1} |$$

It turns out that exponential equality as described above is sufficient to determine the outage probability in the scale of interest. Let

$$\begin{aligned} \mathcal{J} &:= I_m + \mathcal{E}HH^\dagger\Sigma_n^{-1} \\ &= \begin{bmatrix} I_p(1+\rho\gamma_1) & \rho g_1 B^\dagger C^{-1} \\ \rho g_1^* B & I_q + \rho BB^\dagger C^{-1} \end{bmatrix} \end{aligned}$$

where $C = I_q + \sum_{j=2}^{n} |h_j|^2 \frac{\sigma_{v_j}^2}{\sigma_w^2} A_j A_j^\dagger$, $\gamma_1 = |g_1|^2$ and $B$ is as defined in Lemma 1.

Then upon row reduction of $\mathcal{J}$, we obtain

$$\begin{aligned} |\,\mathcal{J}\,| &= |\,(1+\rho\gamma_1)I_p\,| \cdot \;|\, I_q + \frac{\rho}{1+\rho\gamma_1}BB^\dagger C^{-1}\,| \\ &= (1+\rho\gamma_1)^p \,|\, C^{-1}\,|\,|\, C + \frac{\rho}{1+\rho\gamma_1}BB^\dagger\,| \end{aligned} \tag{9}$$

By applying Lemma 1 and using (4), we get

$$\begin{aligned} |\,\mathcal{J}\,| &\;\dot{\leq}\; (1+\rho\gamma_1)^p \cdot \left(1+\sum_{j=2}^{n}\left[|h_j|^2 + \frac{\rho}{1+\rho\gamma_1}\gamma_j\right]\right)^q \\ &\;\dot{\leq}\; (1+\rho\gamma_1)^{p-q}\left(1+\rho\gamma_1+\rho\sum_{j=2}^{n}\gamma_j\right)^q . \end{aligned}$$

As in [5], we define

$$\gamma_1 \;\doteq\; \begin{cases} \rho^{-u}, & u \geq 0 \\ 0, & u < 0 \end{cases}. \tag{10}$$

For $j = 2, 3, \ldots, n$,

$$\gamma_j \;\doteq\; \begin{cases} \rho^{-v_j}, & v_j \geq 0 \\ 0, & v_j < 0 \end{cases} \tag{11}$$

and

$$v = \min\{v_j\}_{j=2}^{n}.$$

This gives us

$$|\,\mathcal{J}\,| \;\dot{\leq}\; \rho^{(p-q)(1-u)^+ + q\max\{(1-u),(1-v)\}^+}.$$

The probability of outage, for the channel in (2), is defined as

$$P_{\text{out}}(R) = \inf_{\Sigma_x \,\geq\, 0,\ \mathrm{Tr}(\Sigma_x) \,\leq\, p\mathcal{E}} \Pr(I(\underline{x};\underline{y}|H) \;\leq\; mR).$$

In the scale of interest, the above expression reduces to (see [2]),

$$P_{\text{out}}(R = r\log\rho) \doteq \Pr(\ \log|\,\mathcal{J}\,| < \ rm\log\rho\ )$$

and is lower bounded by

$$\Pr\left((p-q)(1-u)^+ + q\max\{(1-u),(1-v)\}^+ \;<\; rm\right).$$

Let the negative exponent of $P_{\text{out}}(R)$ be $d(r)$,

$$d(r) \;:=\; -\lim_{\rho\to\infty}\frac{\log P_{\text{out}}(r\log\rho)}{\log\rho}.$$

Then,

$$d(r) \;\leq\; \inf_{(p-q)(1-u)^+ + q\max\{(1-u),(1-v)\}^+ \;<\; mr} u + (n-1)v.$$

It is clear that it is enough to consider $u, v \leq 1$. Hereafter, we will consider $u$ and $v$ to lie in the range $0 \leq u, v \leq 1$. Therefore,

$$d(r) \;\leq\; \inf_{(p-q)u + q\min\{u,v\} \;>\; p-mr} u + (n-1)v.$$

By solving the above optimization problem, we get the statement of the theorem. Please see Appendix II for the details. ■

### C. Specific protocol achieving the DMT upper bound for $p = n,\ q = (n-1)$

Within the class of OAF protocols, there are different protocols corresponding to various choices of $p$, $q$ and $\{A_j\}_{j=2}^{n}$. As seen in Section II-B, for a given number $(n-1)$ of relays, the upper bound on $d(r)$ is maximized when $\frac{p}{p+q} = \frac{n}{2n-1}$.

We have used the inequality

$$\left(\sum_{j=2}^{n} g_j h_j A_j\right)\left(\sum_{j=2}^{n} g_j h_j A_j\right)^{\dagger} \leq (n-1)\left(\sum_{j=2}^{n} \mid g_j h_j \mid^2 A_j A_j^{\dagger}\right) \quad (12)$$

to derive an upper bound on the DMT of the class of OAF protocols. Equality will occur in (12) if

$$A_j A_k^{\dagger} = 0\ , \quad \text{for all } j \neq k, \quad (13)$$

i.e., if the row spaces of the matrices $\{A_j\}_{j=2}^{n}$ are pairwise orthogonal.

The arguments presented above serve as a motivation for the particular choice of the matrices $\{A_j\}$ outlined next. With this specific choice of $\{A_j\}$, satisfying the constraints in (13), we can achieve the upper bound on the DMT given in Theorem 2.

*Theorem 3:* Consider a specific OAF protocol, as described above, with parameters $p = n$ and $q = n-1$. Choose the $(n-1)\times n$ matrices $\{A_j\}$ as follows:

$$A_j(k,l) = \begin{cases} \alpha_j & k = j-1, l = j \\ 0 & \text{elsewhere} \end{cases}, \quad (14)$$

i.e., the $(j-1,j)^{\text{th}}$ entry of $A_j$ is equal to $\alpha_j$ and remaining entries are 0. The DMT of this protocol is equal to the highest upper bound of the class of OAF protocols,

$$d(r) = \begin{cases} n-(2n-1)r & , \quad 0 \leq r \leq \frac{1}{2} \\ 1-r & , \quad \frac{1}{2} < r \leq 1 \end{cases}.$$

*Proof:* With the above choice of $\{A_j\}$ we get,

$$BB^{\dagger} = \begin{bmatrix} \alpha_2^2\gamma_2 & & \\ & \ddots & \\ & & \alpha_n^2\gamma_n \end{bmatrix}.$$

By substituting for $BB^{\dagger}$ in equation (9), and setting $p = n$ and $q = n-1$, we get,

$$\begin{aligned} \mid \mathcal{J} \mid &\doteq (1+\rho\gamma_1)^n . \left|\begin{bmatrix} 1+\frac{\rho}{1+\rho\gamma_1}\gamma_2 & & \\ & \ddots & \\ & & 1+\frac{\rho}{1+\rho\gamma_1}\gamma_n \end{bmatrix}\right| \\ &= (1+\rho\gamma_1)\prod_{j=2}^{n}(1+\rho\gamma_1+\rho\gamma_j) \\ &\doteq \rho^{(1-u)^+}\prod_{j=2}^{n}\rho^{(1-\min\{u,v_j\})^+}. \end{aligned}$$

As in Theorem 2, the outage probability is given by

$$\begin{aligned} &P_{\text{out}}(r\log\rho) \\ &\doteq \Pr(\log \mid \mathcal{J} \mid < (2n-1)r\log\rho) \\ &\doteq \Pr\left((1-u)^+ + (1-\sum_{j=2}^{n}\min\{u,v_j\})^+ < (2n-1)r\right). \end{aligned}$$

Let the negative exponent of $P_{\text{out}}(r\log\rho)$ in the scale of interest be $d(r)$. Then, we can consider $u$ and $v_j$ to lie in the range $0 \leq u, v_j \leq 1$. We get

$$d(r) = \inf_{u+\sum_{j=2}^{n}\min\{u,v_j\} > n-(2n-1)r} \quad u + \sum_{j=2}^{n} v_j .$$

By solving the optimization problem, we get

$$d(r) = n-(2n-1)r\ , \quad 0 \leq r \leq \frac{n}{2n-1}. \quad (15)$$

For $r \geq \frac{1}{2}$, we allow the source to transmit to the destination continuously, thereby achieving the tradeoff mentioned in the statement of the theorem. ■

Since $(n, 2n-1) = 1$, the smallest value of delay parameter $m = (p+q)$ satisfying the condition $\frac{p}{p+q} = \frac{n}{2n-1}$ corresponds to the choice $p = n,\ q = (n-1)$. Hence, the above protocol has minimum possible delay required to achieve the best DMT.

*Example 1:* Let the number of relays be 2. Therefore, $n = 3$. We choose $q = 2$, $p = 3$, and

$$A_2 = \begin{bmatrix} 0 & \alpha_2 & 0 \\ 0 & 0 & 0 \end{bmatrix},\ A_3 = \begin{bmatrix} 0 & 0 & 0 \\ 0 & 0 & \alpha_3 \end{bmatrix}.$$

For these parameters, the DMT of the OAF protocol is

$$d(r) = \begin{cases} 3-5r & , \quad 0 \leq r \leq \frac{1}{2} \\ 1-r & , \quad \frac{1}{2} < r \leq 1 \end{cases}$$

and is shown in Fig.3.

### D. DMT Optimal Codes for the OAF Protocol

Our code construction is based on cyclic division algebras (CDA) and we begin with a brief introduction to these algebraic objects.

*1) Division Algebras:* Division algebras are rings with identity in which every nonzero element has a multiplicative inverse. The center $\mathbb{F}$ of any division algebra $D$, i.e., the subset comprising of all elements in $D$ that commute with every element of $D$, is a field. The division algebra is a vector space over the center $\mathbb{F}$ of dimension $n^2$ for some integer $n$. A field $\mathbb{L}$ such that $\mathbb{F} \subset \mathbb{L} \subset D$ and such that no subfield of $D$ contains $\mathbb{L}$ is called a *maximal subfield* of $D$ (Fig. 4). Every division algebra is also a vector space over a maximal subfield and the dimension of this vector space is the same for all maximal subfields and equal to $n$. This common dimension $n$ is known as the *index* of the division algebra.

$D$ Division Algebra

$n$

$\mathbb{L}$ Maximal Subfield

$n$

$\mathbb{F}$ Centre

Fig. 4. *Structure of a division algebra.*

*2) Cyclic Division Algebras:* Our interest is in CDA, i.e., division algebras in which the center $\mathbb{F}$ and a maximum subfield $\mathbb{L}$ are such that $\mathbb{L}/\mathbb{F}$ is a finite cyclic Galois extension. CDAs have a simple characterization that aids in their construction, see [15, Proposition 11], or [14, Theorem 1].

Let $\mathbb{F}, \mathbb{L}$ be number fields, with $\mathbb{L}$ a finite, cyclic Galois extension of $\mathbb{F}$ of degree $n$. Let $\sigma$ denote the generator of the Galois group $\mathrm{Gal}(\mathbb{L}/\mathbb{F})$. Let $z$ be an indeterminate satisfying

$$\ell z \;=\; z\sigma(\ell) \quad \forall \;\; \ell \in \mathbb{L} \quad \text{and} \quad z^n = \gamma,$$

for some non-norm element $\gamma \in \mathbb{F}^*$, by which we mean some element $\gamma$ having the property that the smallest positive integer $t$ for which $\gamma^t$ is the relative norm $N_{\mathbb{L}/\mathbb{F}}(u)$ of some element $u$ in $\mathbb{L}^*$, is $n$. Then, a CDA $D(\mathbb{L}/\mathbb{F}, \sigma, \gamma)$ with index $n$, center $\mathbb{F}$ and maximal subfield $\mathbb{L}$ is the set of all elements of the form

$$\sum_{i=0}^{n-1} z^i \ell_i, \quad \ell_i \in \mathbb{L}. \tag{16}$$

Moreover, it is known that every CDA has this structure. It can be verified that $D$ is a right vector space (i.e., scalars multiply vectors from the right) over the maximal subfield $\mathbb{L}$.

*3) Space-Time Codes from Cyclic Division Algebras:* A space-time (ST) code $\mathcal{X}$ can be associated to $D$ by selecting the set of matrices corresponding to the matrix representation of elements of a finite subset of $D$. Note that since these matrices are all square matrices, the resultant ST code necessarily has $T = n_t$.

The matrix corresponding to an element $d \in D$ corresponds to left multiplication by the element $d$ in the division algebra. Let $\lambda_d$ denote this operation, $\lambda_d : D \to D$, defined by

$$\lambda_d(e) = de, \; \forall \; e \in D.$$

It can be verified that $\lambda_d$ is a $\mathbb{L}$-linear transformation of $D$. From (16), a natural choice of basis for the right-vector space $D$ over $\mathbb{L}$ is $\{1, z, z^2, \ldots, z^{n-1}\}$. A typical element in the division algebra $D$ is $d = \ell_0 + z\ell_1 + \cdots + z^{n-1}\ell_{n-1}$, where the $\ell_i \in \mathbb{L}$. By considering the effect of multiplying $d \times 1$, $d \times z$, $\ldots$, $d \times z^{n-1}$, one can show that the $\mathbb{L}$-linear transformation $\lambda_d$ under this basis has the matrix representation,

$$\begin{bmatrix} \ell_0 & \gamma\sigma(\ell_{n-1}) & \gamma\sigma^2(\ell_{n-2}) & \ldots & \gamma\sigma^{n-1}(\ell_1) \\ \ell_1 & \sigma(\ell_0) & \gamma\sigma^2(\ell_{n-1}) & \ldots & \gamma\sigma^{n-1}(\ell_2) \\ \vdots & \vdots & \vdots & \ddots & \vdots \\ \ell_{n-1} & \sigma(\ell_{n-2}) & \sigma^2(\ell_{n-3}) & \ldots & \sigma^{n-1}(\ell_0) \end{bmatrix}, \tag{17}$$

known as the left regular representation of $d$.

A set of such matrices, obtained by choosing a finite subset of elements in $D$ constitutes the CDA-based ST code $\mathcal{X}$. In [1], the authors have constructed CDA-based ST code for all values of $n$. For all the codes constructed in [1], the underlying constellation is QAM and the center of the division algebra is $\mathbb{F} = \mathbb{Q}[\imath]$.

*4) DMT optimal CDA-based ST codes for the OAF protocol with parameters $p = n, q = (n-1)$:* In this subsection, we provide an explicit construction of a code, based on CDAs, for the OAF protocol for any number of relays, and prove the DMT optimality of the code. If the number of relays is $n-1$, we choose $p = n$ and $q = (n-1)$ since this choice of parameters has the best DMT (see Theorem 2).

The channel model for the OAF protocol in (2) can be rewritten as

$$\begin{aligned} &\left[\; \underline{y}_1^t \;\; \underline{y}_2^t \;\right] \\ &= \left[\; g_1 \quad g_2h_2 \quad \cdots \quad g_nh_n \;\right] \left[\begin{array}{ccc|c} x_1 & \cdots & x_p & 0 \\ 0 & \cdots & 0 & \underline{x}^t A_2^t \\ \vdots & \ddots & \vdots & \vdots \\ 0 & \cdots & 0 & \underline{x}^t A_n^t \end{array}\right] \\ &\quad + \underline{n}^t \end{aligned} \tag{18}$$

where

$$\underline{n} = \begin{bmatrix} \underline{w}_1 \\ \sum_{j=2}^{n} h_j A_j \underline{v}_j \; + \; \underline{w}_2 \end{bmatrix}.$$

For $M$ even, let $\mathcal{A}_{\text{QAM}}$ denote the $M^2$-QAM constellation given by

$$\mathcal{A}_{\text{QAM}} = \{a + \imath b \;\;|\;\; |a|, |b| \leq M-1, \;\; a, b \text{ odd}\}. \tag{19}$$

Consider a CDA having center $\mathbb{F} \;=\; \mathbb{Q}(\imath)$ and maximum subfield $\mathbb{L}$ that is a degree-$n$ cyclic Galois extension $\mathbb{L}/\mathbb{F}$ of $\mathbb{F}$. Let $\sigma$ be the generator of the cyclic Galois group $\mathrm{Gal}(\mathbb{L}/\mathbb{F})$. Let $\mathcal{O}_{\mathbb{F}}$ and $\mathcal{O}_{\mathbb{L}}$ denote the ring of algebraic integers in $\mathbb{F}$ and $\mathbb{L}$ respectively. It is known that $\mathcal{O}_{\mathbb{F}} \;=\; \mathbb{Z}[\imath]$. Let $\{\beta_1, \ldots, \beta_n\}$ be an integral basis for $\mathcal{O}_{\mathbb{L}}/\mathcal{O}_{\mathbb{F}}$. Let $\mathrm{D}(\mathbb{L}/\mathbb{F}, \sigma, \gamma)$ denote the associated CDA.
Let

$$\ell_i \in \mathcal{A}_{\text{QAM}}(\beta_1, \cdots, \beta_n) \tag{20}$$

where

$$\mathcal{A}_{\text{QAM}}(\beta_1, \ldots, \beta_n) \;=\; \left\{ \sum_i a_i \beta_i \mid a_i \in \mathcal{A}_{\text{QAM}} \right\}.$$

Then, let the signal transmitted by the source in the first

$p = n$ channel uses be given by[2]

$$\underline{x} = \left[ \begin{array}{cccc} \ell_0 & \sigma(\ell_0) & \ldots & \sigma^{n-1}(\ell_0) \end{array} \right]^t \quad (21)$$
$$\text{where } \ell_0 \in \mathcal{A}_{\text{QAM}}(\beta_1, \ldots, \beta_n).$$

We select $\{A_j\}$ as specified in (14). Without loss of generality (insofar as DMT is concerned), for simplicity we set

$$\alpha_j = 1 \;, \quad 2 \leq j \leq n.$$

As we saw earlier in Theorem 3, this choice of $\{A_j\}$ achieves the upper bound on the DMT for the OAF protocol. Hence, the code matrices will be as shown in (22).

The performance of this code is no worse than that obtained by deleting the columns 2 to $n-1$ of all the code matrices $X$. Hence, from here on, we will work with the column-deleted code matrix:

$$X = \left[ \begin{array}{cccc} \ell_0 & & & \\ & \sigma(\ell_0) & & \\ & & \ddots & \\ & & & \sigma^{n-1}(\ell_0) \end{array} \right],$$
$$\text{where } \ell_0 \in \mathcal{A}_{\text{QAM}}(\beta_1, \cdots, \beta_n). \quad (23)$$

*Theorem 4:* The DMT of the above code is

$$d(r) = n - (2n-1)r \;, \quad 0 \leq r \leq \frac{n}{2n-1}.$$

*Proof:* The mentioned tradeoff is the optimal DMT for the OAF protocol in the range $0 \leq r \leq \frac{1}{2}$.

For choice of the $A_j$ matrices given by (14), the covariance matrix of the noise vector $\underline{n}$ is shown in (24). By normalizing the noise variances to unity and noting that $1 + |h_j|^2 \doteq 1$, the covariance matrix becomes identity in the scale of interest. Hence, we can consider the noise to be white in the scale of interest. Also, we impose an energy constraint on the codewords and replace $X$ with $\theta X$, where $\theta$ is chosen to ensure that

$$\|\theta X\|_F^2 \leq (2n-1)\rho.$$

After deleting columns 2 to $n-1$ of the ST code for the OAF protocol, the resultant channel model can be rewritten as

$$\underline{y} = \theta \left[ \begin{array}{cccc} g_1 & & & \\ & g_2 h_2 & & \\ & & \ddots & \\ & & & g_n h_n \end{array} \right] \left[ \begin{array}{c} \ell_0 \\ \sigma(\ell_0) \\ \vdots \\ \sigma^{n-1}(\ell_0) \end{array} \right] + \underline{n}$$
$$\text{where } \ell_0 \in \mathcal{A}_{\text{QAM}}(\beta_1, \cdots, \beta_n).$$

It can be shown that, in the scale of interest, the above channel is equivalent to a parallel channel. Hence, we show that the chosen code is an optimal code for the parallel channel.

To support a data rate of $R_p = r_p \log(\rho)$ on the parallel channel, we need, $M^2 = \rho^{\frac{r_p}{n}}$. The energy requirement forces, $\theta^2 \doteq \rho^{1-\frac{r_p}{n}}$.

[2] $\ell_0$ need not be drawn from a maximal subfield $\mathbb{L}$ of a division algebra. It is enough if $\mathbb{L}$ is an algebraic extension of $\mathbb{Q}[i]$ of degree $n$ and $\ell_0$ is as defined in (21).

Now, the product of the squared norms of the normalized difference code matrices (obtained by scaling the code vectors with $\frac{1}{\sqrt{\rho}}$) is given by

$$\begin{aligned} &\frac{1}{\rho^n} |\theta\ell_0|^2 \cdot |\theta\sigma(\ell_0)|^2 \cdots |\theta\sigma^{n-1}(\ell_0)|^2 \\ &\quad = \frac{\theta^{2n}}{\rho^n} \prod_{i=0}^{n-1} |\sigma^i(\ell_0)|^2 \;\dot{\geq}\; \frac{\rho^{(1-\frac{r_p}{n})n}}{\rho^n} \rho^0 \\ &\quad \dot{\geq} \; \rho^{-r_p}. \end{aligned}$$

Therefore, from [3, Theorem 5.1] the code is DMT optimal for the parallel channel. It can be shown that the DMT of the parallel channel is (see [3])

$$d_p(r_p) = n - r_p, \;, \quad 0 \leq r_p \leq n. \quad (25)$$

Let the rate of the original code for the OAF protocol be $R = r \log \rho$. Then, the size of the DMT optimal code book for the parallel channel is $\rho^{r_p}$. This corresponds to a code book for the OAF channel of size $\rho^{(2n-1)r}$. Since the two code books are the same, it follows that $r_p = (2n-1)r$. By substituting for $r_p$ in (25), we obtain a lower bound on the DMT of the code for the OAF protocol. The lower bound occurs because dropping some columns of the code matrices could conceivably decrease the probability of error. Therefore,

$$\begin{aligned} d(r) &\geq d_p((2n-1)r) \\ &\geq n - (2n-1)r \;, \quad 0 \leq r \leq \frac{n}{2n-1}. \end{aligned}$$

The above bound equals the DMT of the OAF protocol in (15). Therefore, the DMT of the code is as mentioned in the theorem.

For $r > \frac{1}{2}$, the source will transmit continuously to the destination and the relays will not participate. It follows that the code is DMT optimal for the class of OAF protocols considered here. ■

*Remarks:* We make the following remarks on the class of OAF protocols and the proposed DMT optimal code:

1) Among the class of OAF protocols, the best DMT is achieved when $\frac{p}{m} = \frac{n}{2n-1}$. Since $n$ and $2n-1$ are relatively prime, the proposed code has the minimum possible delay of $2n-1$ with the parameters $p = n$ and $q = n-1$.
2) For $p = n$ and $q = n-1$, the DMT of the OAF protocol, with the choice of $\{A_j\}$ mentioned in Theorem 3 coincides with the DMT of the NAF protocol [5].
3) When each node in the system has only one transmit and one receive antenna, the DMT optimal code for the NAF protocol proposed in [9] has delay $4(n-1)$ which is larger than the delay for the DMT optimal code for the OAF protocol.

## III. Non-Orthogonal Selection Decode and Forward Protocol

In this section, we consider a non-orthogonal selection decode and forward (NSDF) protocol in which the source transmits a signal to the destination and the relays for $p$ channel uses in the first phase. All the relays, which are not

$$X = \left[ \begin{array}{cccc|ccc} \ell_0 & \sigma(\ell_0) & \cdots & \sigma^{n-1}(\ell_0) & 0 & \cdots & 0 \\ 0 & 0 & \cdots & 0 & \sigma(\ell_0) & & \\ \vdots & \ddots & & \vdots & & \ddots & \\ 0 & & \ddots & 0 & & & \sigma^{n-1}(\ell_0) \end{array} \right], \quad \text{where } \ell_0 \in \mathcal{A}_{\text{QAM}}(\beta_1, \cdots, \beta_n). \tag{22}$$

$$\Sigma_n = \begin{bmatrix} \sigma_w^2 & & & & & \\ & \ddots & & & & \\ & & \sigma_w^2 & & & \\ & & & \sigma_w^2 + \sigma_{v_2}^2 |h_2|^2 & & \\ & & & & \ddots & \\ & & & & & \sigma_w^2 + \sigma_{v_n}^2 |h_n|^2 \end{bmatrix}. \tag{24}$$

in outage[3], will decode the source message. In the second phase, the relays will separately encode and transmit a vector of length $q$. The source continues to transmit to the destination in the second phase. We only consider the case when $p \geq q$. To compute the best possible DMT, we allow $p$ and $q$ to vary with the multiplexing gain $r$ and choose the value of $\kappa = \frac{p}{q}$ which maximizes the DMT for a given $r$. This version of the protocol will be called the variable-NSDF protocol. We have also computed the DMT of the fixed-NSDF protocol, wherein the ratio $\kappa = \frac{p}{q}$ is fixed for all $r$. We have constructed CDA based codes which achieve the DMT of the variable and fixed NSDF protocols.

### A. DMT of NSDF Protocol

*Theorem 5:* The DMT of the variable-NSDF protocol is given by

$$d(r) = \begin{cases} n\left(1 - \frac{(n-1)(\kappa_n+1)}{n} r\right), & 0 \leq r \leq \frac{1}{\kappa_n+1} \\ \frac{(n-r)(1-r)}{(n-2)r+1}, & \frac{1}{\kappa_n+1} \leq r \leq 1 \end{cases}, \tag{26}$$

where $\kappa_n = \frac{1+\sqrt{1+4(n-1)^2}}{2(n-1)}$.

In deriving the above DMT, we have allowed $p$ and $q$ to vary with the multiplexing gain $r$. The source and the relays choose a code corresponding to each $p$ and $q$. For the case $p \geq q$, we select the value of $(p, q)$ which maximizes the DMT for a given $r$. Suppose $p = \kappa q$, then the optimal value of $\kappa$ is given by

$$\kappa = \begin{cases} \kappa_n, & 0 \leq r \leq \frac{1}{\kappa_n+1} \\ \frac{1+(n-2)r}{(n-1)(1-r)}, & \frac{1}{\kappa_n+1} < r \leq 1 \end{cases}.$$

For a fixed choice $\kappa = \frac{p}{q}$, the DMT of the fixed-NSDF protocol is given by:

if $1 \leq \kappa \leq \kappa_n$,

$$d(r) = (n-1)\left(1 - \frac{mr}{p}\right)^+ + (1-r), \quad 0 \leq r \leq 1, \tag{27}$$

[3]We say that a relay is not in outage if the corresponding source-relay channel is not in outage.

else if $\kappa \geq \kappa_n$,

$$d(r) = \begin{cases} n\left(1 - \frac{m(n-1)}{nq} r\right), & 0 \leq r \leq \frac{q}{m} \\ \frac{m}{p}(1-r), & \frac{q}{m} \leq r \leq \frac{np-m}{(n-2)m+p} \\ n\left(1 - \frac{(n-1)m+p}{np} r\right), & \frac{np-m}{(n-2)m+p} \leq r \leq \frac{p}{m} \\ 1-r & \frac{p}{m} \leq r \leq 1 \end{cases}. \tag{28}$$

*Proof:* Consider the system model described in Section I. Let $\underline{x_1}$ and $\underline{x_1'}$ be the signals transmitted by the source in the first and second phase respectively. Let $(k-1)$ relays, where $1 \leq k \leq n$, participate in the second phase and let $\{\underline{x_j}\}_{j=2}^k$ be the signal transmitted by them. Only the relays that are not in outage in the first phase shall participate in the cooperative protocol in the second phase.

Let $E_k$ denote the event when any $(k-1)$ relays participate in the cooperative (second) phase. The events $\{E_1, E_2, \ldots, E_n\}$ are disjoint and their probabilities sum up to 1. First, we will calculate the outage probability of the channel when the event $E_k$ occurs.

*1) Outage Probability conditioned on $E_k$:* Consider the case when the event $E_k$, $2 \leq k \leq n$, occurs. The case $k = 1$ will be dealt with separately. Let the signals received by the destination in the two phases be $\underline{y_1}$ and $\underline{y_2}$, where

$$\underline{y_1} = g_1 \underline{x_1} + \underline{w_1}$$

and

$$\underline{y_2} = g_1 \underline{x_1'} + \sum_{j=2}^{k} h_j \underline{x_j} + \underline{w_2}$$

with $\underline{w_1}$, $\underline{w_2}$ denoting the noise added at the destination in the respective phases. We impose an energy constraint by choosing $\mathcal{E}$ to be the average energy available for transmission of a symbol at either the source or a relay. Let $\sigma_w^2$ be the variance of the noise added at the destination and SNR be as defined in (3). We shall normalize the noise variances to unity. Hence, we can regard $\mathcal{E}$ as the SNR $\rho$.

The channel model for the NSDF protocol is given in (29).

$$\begin{bmatrix} \underline{y_1} \\ \underline{y_2} \end{bmatrix} = \begin{bmatrix} g_1 I_p & 0 & 0 & \cdots & 0 \\ 0 & g_1 I_q & h_2 I_q & \cdots & h_k I_q \end{bmatrix} \begin{bmatrix} \underline{x_1} \\ \underline{x_1}' \\ \underline{x_2} \\ \vdots \\ \underline{x_k} \end{bmatrix} + \begin{bmatrix} \underline{w_1} \\ \underline{w_2} \end{bmatrix} \tag{29}$$

Let,

$$\begin{aligned} \underline{y} &= \begin{bmatrix} \underline{y_1} \\ \underline{y_2} \end{bmatrix}, \quad \underline{w} = \begin{bmatrix} \underline{w_1} \\ \underline{w_2} \end{bmatrix}, \\ \underline{x}^t &= \begin{bmatrix} \underline{x_1}^t & \underline{x_1}'^t & \underline{x_2}^t & \cdots & \underline{x_k}^t \end{bmatrix} \end{aligned}$$

and

$$H_k = \begin{bmatrix} g_1 I_p & 0 & 0 & \cdots & 0 \\ 0 & g_1 I_q & h_2 I_q & \cdots & h_k I_q \end{bmatrix}.$$

Then, (29) can be rewritten as

$$\underline{y} = H_k \underline{x} + \underline{w}.$$

The input vector can be assumed to be circularly symmetric complex gaussian. The maximum mutual information transferred between the source and the destination, conditioned on the knowledge of $H_k$ at the destination, is given by

$$I(\underline{y}; \underline{x} | H_k) = \log \mid I_m + H_k \Sigma_x H_k^\dagger \mid .$$

Arguing as in case of the OAF protocol, in the scale of interest, the maximum mutual information between $\underline{x}$ and $\underline{y}$ is given by

$$\begin{aligned} \mathcal{I}_{\max} &\doteq \log \mid I_m + \rho H_k H_k^\dagger \mid \\ &= \log \left(1 + \rho |g_1|^2\right)^p (1 + \rho |g_1|^2 + \rho \sum_{j=2}^{k} |h_j|^2)^q \\ &\doteq \log \rho^{p(1-u)^+ + q \max\{(1-u),(1-v)\}^+} \end{aligned}$$

where,

$$\begin{aligned} |g_1|^2 &\doteq \begin{cases} \rho^{-u}, & u \geq 0 \\ 0, & u < 0 \end{cases}, \\ |h_j|^2 &\doteq \begin{cases} \rho^{-v_j}, & v_j \geq 0 \\ 0, & v_j < 0 \end{cases}, \quad j = 2, \ldots, k \end{aligned}$$

and

$$v = \min \{v_j\}_{j=2}^k.$$

The outage probability of the channel in (29), in the scale of interest, is given by,

$$P_{\text{out}}(r \log \rho) = \Pr\left(\mathcal{I}_{\max} < mr \log \rho\right) := \rho^{-d_k(r)},$$

so that,

$$d_k(r) = \inf_{p(1-u)^+ + q \max\{(1-u),(1-v)\}^+ < mr} u + (k-1)v. \tag{30}$$

It is clear that it is enough to consider $0 \leq u, v \leq 1$. Hence, the above infimum must be calculated over the region

$$pu + q \min\{u, v\} > m(1-r), \quad 0 \leq u, v \leq 1.$$

It is clear that for $r > 1$, $d_k(r) = 0$. Therefore, it is enough to consider $r \leq 1$. We consider two separate cases to evaluate $d_k(r)$. We set $p \geq q$ in the optimization that follows.

Case I: $\min\{u, v\} = u$

We get,

$$u > (1-r) \text{ and } v > u.$$

Substituting in (30) we get,

$$d_k(r) = k(1-r), \quad 0 \leq r \leq 1.$$

Case II: $\min\{u, v\} = v$

We have,

$$pu + qv > m(1-r).$$

As in the case of the OAF protocol, we solve the optimization problem to get,

$$d_k(r) = \begin{cases} k - \frac{(k-1)m}{q} r, & 0 \leq r \leq \frac{q}{m} \\ \frac{m}{p}(1-r), & \frac{q}{m} \leq r \leq 1 \end{cases}. \tag{31}$$

Now, we handle the case when event $E_1$ occurs, i.e., $k = 1$. The channel model in this case is given by

$$\begin{aligned} \begin{bmatrix} \underline{y_1} \\ \underline{y_2} \end{bmatrix} &= \begin{bmatrix} g_1 I_p & 0 \\ 0 & g_1 I_q \end{bmatrix} \begin{bmatrix} \underline{x_1} \\ \underline{x_1}' \end{bmatrix} + \begin{bmatrix} \underline{w_1} \\ \underline{w_2} \end{bmatrix} \\ &= H_1 \underline{x} + \underline{w}. \end{aligned} \tag{32}$$

The probability of outage of the above channel is given by

$$P_{\text{out}}(r \log \rho) := \rho^{-d_1(r)} \doteq \rho^{-(1-r)^+}. \tag{33}$$

*2) Probability of the set of Participating Relays:* In this subsection, we will compute the probability of the event $E_k$. The probability of $E_k$ is the product of probabilities of two events:

$$\begin{aligned} \Pr(E_k) = \; & \Pr((n-k) \text{ relays are in outage}) \cdot \\ & \Pr((k-1) \text{ relays are not in outage}). \end{aligned}$$

We shall evaluate the probabilities of the two events mentioned above separately. Also, we say that a relay is participating in the second phase if the corresponding source-relay channel is not in outage.

The signal received by the $j$th relay $R_j$, where $2 \leq j \leq n$, in the first phase is given by

$$\underline{r_j} = g_j \underline{x_1} + \underline{v_j},$$

where $v_j$ is the noise vector. The maximum mutual information between $\underline{x_1}$ and $\underline{r_j}$, conditioned on the knowledge of $g_j$ at the relay, is

$$\begin{aligned} \mathcal{I}_{\max} &= \max_{\Sigma_{x_1} \geq 0, \; \text{Tr}(\Sigma_{x_1}) \leq p\mathcal{E}} I(\underline{r_j}; \underline{x_1} | g_j) \\ &\doteq \frac{p}{m} \log(1 + \rho |g_j|^2). \end{aligned}$$

Now, the probability that $R_j$ is in outage is

$$\begin{aligned}\Pr(R_j \text{ is in outage}) &\doteq \Pr\left(\frac{p}{m}\log(1+\rho|g_j|^2) < r\log\rho\right)\\ &\doteq \Pr\left((1-u_j)^+ < \frac{mr}{p}\right)\\ &\doteq \rho^{-(1-\frac{mr}{p})^+}\end{aligned}$$

where

$$|g_j|^2 \doteq \begin{cases} \rho^{-u_j}, & u_j \geq 0 \\ 0, & u_j < 0 \end{cases}.$$

Since the fading coefficients corresponding to different source-relay channels are independent of each other, we have

$$\Pr\left((n-k) \text{ relays are in outage}\right) \doteq \rho^{-(n-k)(1-\frac{mr}{p})^+}.$$

The probability of a relay participating in the second phase will be determined separately for $r \leq \frac{p}{m}$ and $r > \frac{p}{m}$. When $0 \leq r \leq \frac{p}{m}$,

$$\begin{aligned}&\Pr\left((k-1) \text{ relays participate}\right)\\ &\quad= \left(1-\Pr\left(\text{a relay is in outage}\right)\right)^{k-1}\\ &\quad\doteq \left(1-\rho^{-(1-\frac{mr}{p})^+}\right)^{k-1}\\ &\quad\doteq 1.\end{aligned}$$

When $r > \frac{p}{m}$, the probability that a particular relay participates is given by

$$\begin{aligned}&\Pr\left(R_j \text{ participates in second phase}\right)\\ &\doteq \Pr\left(\frac{p}{m}\log(1+\rho|g_j|^2) > \left(\frac{p}{m}+\epsilon\right)\log\rho\right)\\ &= \Pr\left(\log(1+\rho|g_j|^2) > \left(1+\frac{m}{p}\epsilon\right)\log\rho\right)\\ &\doteq \Pr\left(|g_j|^2 > \rho^{\frac{m}{p}\epsilon}\right)\\ &\doteq \rho^{-\infty}\\ &\doteq 0.\end{aligned}$$

Therefore, when $r > \frac{p}{m}$,

$$\Pr\left(k \text{ relays participate in second phase}\right) \doteq 0, \quad 2 \leq k \leq n.$$

Hence, when $r > \frac{p}{m}$, all the relays are in outage with probability 1.

Consolidating the above facts, we get

$$\Pr(E_k) \doteq \begin{cases} \rho^{-(n-k)\left(1-\frac{mr}{p}\right)^+}, & 0 \leq r \leq \frac{p}{m} \\ 0, & r > \frac{p}{m},\ 2 \leq k \leq n \\ 1, & r > \frac{p}{m},\ k=1 \end{cases}. \quad (34)$$

*3) Outage Probability of the NSDF Protocol:* The probability of outage of the NSDF protocol can be calculated as follows:

$$P_{out}(R) = \sum_{k=1}^{n} \binom{n-1}{k-1} \Pr(E_k)\ \Pr(H_k \text{ in outage}|E_k) \quad (35)$$

Let $P_{out}(r\log\rho) := \rho^{-d(r)}$.

It follows from (34) that for $r > \frac{p}{m}$ no relay will participate in the second phase and the channel will be as shown in (32). Hence, from (33), we can see that,

$$d(r) = (1-r)^+, \quad \frac{p}{m} < r \leq 1. \quad (36)$$

For $0 \leq r \leq \frac{p}{m}$, by substituting (31) and (34) in (35), we get

$$\begin{aligned}d(r) = \min_{2\leq k\leq n}\Big\{&(n-1)\left(1-\frac{mr}{p}\right)+(1-r),\\ &(n-k)\left(1-\frac{mr}{p}\right)+d_k(r)\Big\}.\end{aligned}$$

By solving the optimization problem, we get the statement in Theorem 5. Please see Appendix III for the details. ∎

### *B. DMT Optimal Codes for NSDF Protocol:*

In this subsection, we construct a DMT optimal code for the NSDF protocol. Once again, we shall use CDA's to construct ST codes and derive a code for the NSDF protocol from the set of matrices comprising the ST code. In Theorem 5, we have allowed $p$ and $q$ to vary with the multiplexing gain $r$. Now, for a fixed $\kappa = \frac{p}{q}$, where $p$ and $q$ are relatively prime, we outline the construction of a ST code when $(n-1)$ relays are employed in a cooperative network[4].

Let $t = p + nq$. Consider the CDA $\mathrm{D}(\mathbb{L}/\mathbb{F}, \sigma, \gamma)$, as described in Section II-D.2, with the maximum subfield $\mathbb{L}$ being a degree $t$ cyclic Galois extension of the field $\mathbb{F}$. Consider the space time code $\mathcal{X}$ comprising of matrices corresponding to the left-regular representation, as in (17)[5], of all the elements in the CDA D. Let $\mathcal{Z}$ denote the normalized code

$$\mathcal{Z} = \{\theta X \mid X \in \mathcal{X}\}$$

where $\theta$ is chosen to ensure that

$$\|\theta X\|_F^2 \leq t\rho, \quad \text{for all} \quad X \in \mathcal{X}.$$

The transmitted code matrix, denoted by $Z$, will be of the form

$$\theta \begin{bmatrix} \ell_0 & \gamma\sigma(\ell_{t-1}) & \gamma\sigma^2(\ell_{t-2}) & \dots & \gamma\sigma^{t-1}(\ell_1) \\ \ell_1 & \sigma(\ell_0) & \gamma\sigma^2(\ell_{t-1}) & \dots & \gamma\sigma^{t-1}(\ell_2) \\ \ell_2 & \sigma(\ell_1) & \sigma^2(\ell_0) & \dots & \gamma\sigma^{t-1}(\ell_3) \\ \vdots & \vdots & \vdots & \ddots & \vdots \\ \ell_{t-1} & \sigma(\ell_{t-2}) & \sigma^2(\ell_{t-3}) & \dots & \sigma^{t-1}(\ell_0) \end{bmatrix},$$

where $\ell_i$ are as defined in (20). In the first phase the source transmits the first $p$ rows of $Z$ and in the second phase the source transmits the next $q$ rows of $Z$. The $j^{\text{th}}$ relay $R_j$, $2 \leq j \leq n$, if not in outage, will decode the $p$ rows transmitted by the source and will transmit $q$ rows of $Z$ numbering from $p+(j-1)q+1$ to $p+jq$. Total delay of the DMT optimal code will be $(p+q)(p+nq)$ time slots.

*Theorem 6:* The ST code $\mathcal{Z}$ described above achieves the DMT of the fixed-NSDF protocol.

[4]We considering only rational values of $\kappa$ here although, while computing the DMT, $\kappa$ was allowed to take on irrational values too.

[5]The parameter $n$ in (17) and in this section refer to two distinct entities.

*Proof:* We will prove that the given code is DMT optimal irrespective of the number of participating relays.

From [1] and [3], it follows that the given code is approximately universal and hence, DMT optimal for any channel. Also, from [1, Theorem 12], we know that the code $\mathcal{Z}$ remains approximately universal even if we remove a particular set of rows from all the matrices in the code.

Now, during the first phase, the source transmits the first $p$ rows of $Z$ and this code will be approximately universal for the channel seen by the $j^{\text{th}}$ relay. So, when $R_j$ is not in outage, it will be able to decode the source's transmission and participate in the second phase. Consider the case when $k-1$ relays, where $2 \leq k \leq n$, are participating in the second phase. The channel model will be as shown in (29), with $H_k$ being a $m \times (p+kq)$ matrix. Since each node (either source or relay) transmits only $q$ rows of the code matrix $Z$, the code for the channel in (29) corresponds to deletion of $(n-k)q$ rows from $Z$. Therefore, the row-deleted code is DMT optimal for the channel in (29) for all $k = 2, 3, \ldots, n$. When no relays participate in the second phase, the channel model is as shown in (32). Once again, the row-deleted code is DMT optimal for this channel. Hence, the given code is DMT optimal for the NSDF protocol. ■

By constructing a DMT optimal code for each value of the ratio $\kappa = \frac{p}{q}$, we can construct DMT optimal codes for the variable-NSDF protocol.

*Remarks:* We mention the salient features of the results in this section below.

- In the class of static AF and DF protocols, the variable-NSDF protocol has the best DMT for any number of relays, except for the case of two relays.
- The DMT of the variable-NSDF protocol for the case of two relays is better than the tradeoff of the SAF protocol [10] for $r > 0.6$ (see Fig. 3).
- For $\kappa$ in the range $1 < \kappa < \frac{n}{n-1}$, the fixed-NSDF protocol has a better DMT than that of the NAF protocol for any number of relays.
- For $\kappa = 1$, the fixed-NSDF protocol and the NAF protocol have the same DMT.
- When $p = q = 1$, the DMT of the fixed-NSDF protocol coincides with that of the NAF protocol. However, the DMT optimal code for the fixed-NSDF protocol has a delay $2(n+1)$, where $(n-1)$ is the total number of relays, which is considerably shorter than the delay for the DMT optimal codes for the NAF protocol constructed in [9] for $n \geq 3$. The codes in [9] have delay $4(n-1)$.
- Surprisingly, for the case of one relay the optimal ratio $\kappa_2$ turns out to be the Golden Number, $\kappa_2 = \frac{1+\sqrt{5}}{2}$!

## IV. Orthogonal Selection Decode and Forward Protocol

In this section, we consider the orthogonal selection decode and forward (OSDF) protocol. The OSDF protocol is the same as the NSDF protocol, except that the source remains silent in the second phase.

We state the DMT of the variable-OSDF and fixed-OSDF protocol, but we omit the proof since the DMT can be obtained along similar lines to the derivation of the DMT for the NSDF protocol. Also, an approximately universal CDA code of dimension $(p+(n-1)q) \times (p+(n-1)q)$, where $(n-1)$ is the number of relays, will be DMT optimal for the OSDF protocol. The transmission of various rows of the code matrices by the source and relay will be on similar lines to that mentioned in Section III-B for the NSDF protocol.

### A. DMT of OSDF Protocol

*Theorem 7:* The DMT of the variable-OSDF protocol is given by

$$d(r) = \begin{cases} n\left(1 - \frac{(n-1)(\kappa_n+1)}{n} r\right), & 0 \leq r \leq \frac{1}{\kappa_n+1} \\ \frac{n(1-r)}{(n-1)r+1}, & \frac{1}{\kappa_n+1} \leq r \leq 1 \end{cases},$$

where $\kappa_n = \frac{n}{n-1}$.

As in the variable-NSDF protocol, we have allowed $p$ and $q$ to vary with the multiplexing gain $r$. We choose the value of $(p, q)$ which maximizes the diversity, for a given $r$, for the case $p \geq q$. Suppose $p = \kappa q$, then the optimal value of $\kappa$ is given by

$$\kappa = \begin{cases} \kappa_n, & 0 \leq r \leq \frac{1}{\kappa_n+1} \\ \frac{1+(n-1)r}{(n-1)(1-r)}, & \frac{1}{\kappa_n+1} < r \leq 1 \end{cases}.$$

For a fixed choice $\kappa = \frac{p}{q}$, the DMT of the fixed-OSDF protocol is given by:

if $1 \leq \kappa \leq \kappa_n$,

$$d(r) = \begin{cases} n\left(1 - \frac{mr}{p}\right), & 0 \leq r \leq \frac{(n-1)p}{nm-p} \\ (1-r), & \frac{(n-1)p}{nm-p} \leq r \leq 1 \end{cases}$$

else if $\kappa \geq \kappa_n$,

$$d(r) = \begin{cases} n\left(1 - \frac{m(n-1)}{nq} r\right), & 0 \leq r \leq \frac{q}{m} \\ \frac{m}{p}(1-r), & \frac{q}{m} \leq r \leq \frac{np-m}{m(n-1)} \\ n\left(1 - \frac{mr}{p}\right), & \frac{np-m}{m(n-1)} \leq r \leq \frac{(n-1)p}{nm-p} \\ (1-r), & \frac{(n-1)p}{nm-p} \leq r \leq 1 \end{cases}.$$

## V. Non Orthogonal Amplify and Forward Protocol

In this section, we construct a code that is DMT optimal for the non-orthogonal amplify-and-forward (NAF) protocol. For the sake of completeness, we will reproduce here the description of the NAF protocol. The DMT of the protocol was first computed in [5].

In this protocol, the source $S$ transmits at each time instant, and the relays take turns in transmitting an amplified version of a previously received signal. If the number of relays is $(n-1)$, the set of equations describing a $2(n-1)$-length frame are (see [5])

$$\begin{aligned} y_t &= g_1 x_t + w_t, & & t \text{ odd} \\ y_t &= b_i h_i (g_i x_{t-1} + v_i) + g_1 x_t + w_t, & & \begin{cases} i = \frac{t}{2}+1 \\ t \text{ even} \end{cases}, \end{aligned} \tag{37}$$

where $b_i$ is the amplification factor at relay $R_i$, $x_t$ is the signal transmitted at time instant $t$, $y_t$ is the signal received at the destination at time $t$ and $w_t$ is the noise added at the destination at time $t$. $v_i$ is the noise added at the relay $R_i$ in this frame.

*Theorem 8:* [5, Theorem 4] The DMT of the NAF protocol is given by

$$d(r) = (1-r)^+ + (n-1)(1-2r)^+.$$

### A. Explicit DMT optimal codes for NAF protocol

We will present an explicit construction, based on CDA, which achieves the DMT of the NAF protocol.

Consider a $2(n-1)\times 2(n-1)$ DMT optimal CDA ST code. In accordance with the NAF protocol in [5], let the source continuously transmit the vector $[x_1, x_2, \cdots, x_{4(n-1)^2}]$, coming from a row by row vectorization of the code. Each intermediate relay $R_i$, $i = 2, \cdots, n$, forwards at time $t = 4(n-1)(i-2) + 2(n-1) + k$ what it received at time $t = 4(n-1)(i-2)+k$ where $k = 1, 2, \cdots, 2(n-1)$ .

*Theorem 9:* The above scheme achieves the DMT of the NAF protocol.

*Proof:* For the single relay case, we use the equivalent representation of the channel for the NAF protocol in matrix form:

$$\begin{bmatrix} y_1 & y_3 \\ y_2 & y_4 \end{bmatrix} = \begin{bmatrix} g_1 & 0 \\ g_2h_2 & g_1 \end{bmatrix} \begin{bmatrix} x_1 & x_3 \\ x_2 & x_4 \end{bmatrix} + \begin{bmatrix} w_1 & w_3 \\ h_2v_{2,1}+w_2 & h_2v_{2,3}+w_4 \end{bmatrix}.$$

On vectorizing the above channel, we can see that the noise vector is white in the scale of interest. Hence, the DMT of the above channel is met by the approximately universal $2\times 2$ CDA code.

Proceeding as in the single relay case, we can show that the DMT of the NAF protocol is achieved by the corresponding approximately universal $2(n-1)\times 2(n-1)$ CDA code. ■

The above code for the NAF protocol was first presented in [17]. Around the same time, in [9], the authors constructed DMT optimal codes for the NAF protocol which have shorter delays than the codes presented here.

## Appendix I
## Proof of Lemma 1

*Proof:* Let $f_j = g_jh_j$. Then $B = \sum_{j=2}^n f_jA_j$. We have,

$$\begin{aligned} \underline{x}^\dagger BB^\dagger \underline{x} &= \|B^\dagger \underline{x}\|^2 \\ &= \left\| \sum_{j=2}^n f_j^* A_j^\dagger \underline{x} \right\|^2 . \end{aligned}$$

Let $f_j^* A_j^\dagger \underline{x} = \underline{y}_j = (y_{i1}\ y_{i2}\ \cdots\ y_{ip})$. Hence,

$$\begin{aligned} \underline{x}^\dagger BB^\dagger \underline{x} &= \left\| \sum_{j=2}^n \underline{y}_j \right\|^2 \\ &= \sum_{i=1}^p \left| \sum_{j=2}^n y_{ij} \right|^2 . \end{aligned}$$

By applying the Cauchy-Schwarz inequality, we get

$$\begin{aligned} \underline{x}^\dagger BB^\dagger \underline{x} &\le \sum_{i=1}^p \left( \sum_{j=2}^n |y_{ij}|^2 \right) \left( \sum_{j=2}^n 1^2 \right) \\ &= (n-1) \sum_{i=1}^p \sum_{j=2}^n |y_{ij}|^2 \\ &= (n-1) \sum_{j=2}^n \|y_j\|^2 \\ &= (n-1) \sum_{j=2}^n \|f_j^* A_j^\dagger \underline{x}\|^2 \\ &= (n-1) \underline{x}^\dagger \sum_{j=2}^n |f_j|^2 A_j A_j^\dagger \underline{x}. \end{aligned}$$

With this, we get the bound stated in the Lemma. ■

## Appendix II
## Solution of the Optimization problem for the OAF Protocol

We need to solve

$$d(r) \le \inf_{(p-q)u + q\min\{u,v\} \,>\, p-mr} u + (n-1)v \ . \quad (38)$$

If $p = m$, we get the non-cooperative case with the DMT given by

$$d(r) = (1-r)^+.$$

Now, we separately consider different ranges of $r$. The optimization procedure is summarized in the flowchart in Fig. 5.

<u>Case A</u> $r > \frac{p}{m}$

Since $r > \frac{p}{m}$, $p - mr$ is negative. Hence, the infimum is obtained by choosing $u = v = 0$ and

$$d\,(r) \le 0.$$

But, by definition, $d(r)$ cannot be negative. Therefore,

$$d(r) = 0\ , \quad \text{for } r > \frac{p}{m}.$$

<u>Case B</u> $0 \le r \le \frac{p}{m}$

As shown in Fig.5, this case further breaks up into two cases.

<u>Case B.I</u> $\min\{u, v\} = u$

Consider the inequality

$$pu > p - mr,$$

which leads to

$$u > 1 - \frac{mr}{p}.$$

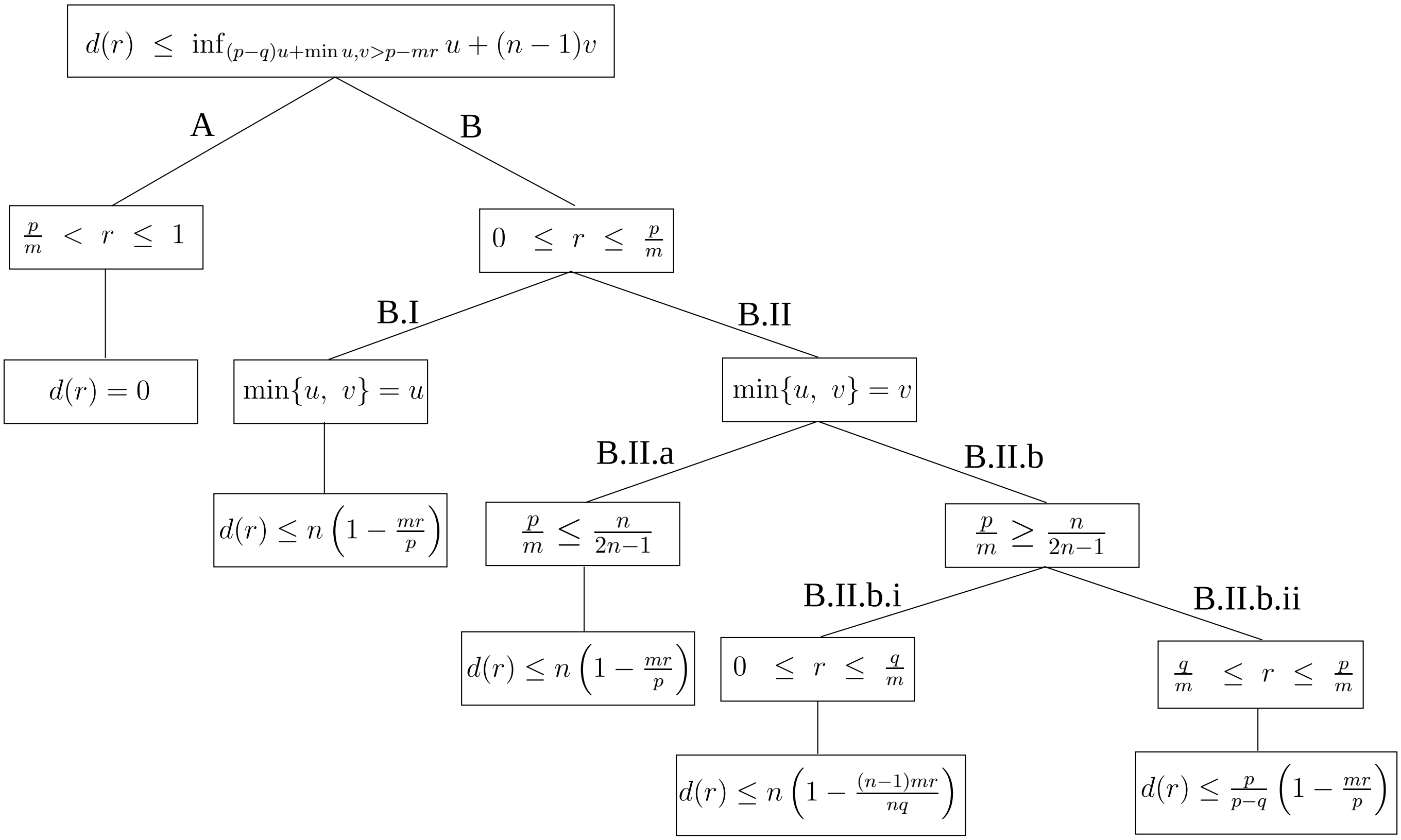

Fig. 5. *Flowchart describing the computation of the DMT of the OAF protocol.*

Therefore, by substituting in (38), we get

$$d(r) \quad \le \quad n\left(1-\frac{mr}{p}\right).$$

Case B.II $\min\{u,v\}=v$
One solution to the optimization problem in (38) is

$$u=v=1-\frac{mr}{p}.$$

We can choose

$$\begin{aligned} u &= 1-\frac{mr}{p}+\delta \\ v &= 1-\frac{mr}{p}-\delta\left(\frac{p-q}{q}\right), \end{aligned}$$

where $\delta$ is a small positive number. In this case,

$$\begin{aligned} & u+(n-1)v \\ &= 1-\frac{mr}{p}+\delta+(n-1)\left(1-\frac{mr}{p}-\delta\frac{p-q}{q}\right) \\ &= n\left(1-\frac{mr}{p}\right)+\delta\left(1-(n-1)\,\frac{p-q}{q}\right). \end{aligned} \tag{39}$$

Now, we need to consider two separate cases depending on the ratio $\frac{p}{m}$.
Case B.II.a $\frac{p}{m} \le \frac{n}{2n-1}$
Suppose

$$1-(n-1)\,\frac{p-q}{q} \;\ge\; 0,$$

we get

$$\frac{p}{m} \;\le\; \frac{n}{2n-1}.$$

In this case, we can substitute $\delta=0$ in (39) to obtain the infimum. Hence,

$$d(r) \quad \le \quad n\left(1-\frac{mr}{p}\right).$$

Case B.II.b $\frac{p}{m} \ge \frac{n}{2n-1}$
In this case, we choose $\delta$ as large as possible under the constraints $u \le 1$ and $v \ge 0$. Suppose $u \le 1$,

$$\begin{aligned} 1-\frac{mr}{p}+\delta &\le 1 \\ \delta &\le \frac{mr}{p}. \end{aligned} \tag{40}$$

Suppose $v \;\ge\; 0$, we have

$$\begin{aligned} 1-\frac{mr}{p}-\delta\left(\frac{p-q}{q}\right) &\ge 0 \\ \delta &\le \frac{q}{p-q}\left(1-\frac{mr}{p}\right). \end{aligned} \tag{41}$$

Hence, $\delta$ is chosen to meet one of the upper bounds in (40) or (41) depending on $r$. By equating the two upper bounds, we get

$$r \;=\; \frac{q}{m}.$$

So, we need to consider two different ranges of $r$.

<u>Case B.II.b.i</u> $0 \leq r \leq \frac{q}{m}$
We choose
$$\delta = \frac{mr}{p}.$$

Then,
$$\begin{aligned} d(r) &\leq n\left(1-\frac{mr}{p}\right)+ \\ &\qquad \frac{mr}{p}\left[1-(n-1)\left(\frac{p-q}{q}\right)\right] \\ &\leq n\left(1-\frac{(n-1)mr}{nq}\right). \end{aligned}$$

<u>Case B.II.b.ii</u> $\frac{q}{m} \leq r \leq \frac{p}{m}$
We choose
$$\delta = \left(\frac{q}{p-q}\right)\left(1-\frac{mr}{p}\right).$$

Then,
$$\begin{aligned} d(r) &\leq n\left(1-\frac{mr}{p}\right)+ \\ &\quad \left(\frac{q}{p-q}\right)\left(1-\frac{mr}{p}\right)\left[1-(n-1)\left(\frac{p-q}{q}\right)\right] \\ &\leq \frac{p}{p-q}\left(1-\frac{mr}{p}\right). \end{aligned}$$

<u>*Summarizing the various cases*</u>
We make some comments with respect to the flowchart in Fig. 5.

- From Case A, it follows that
$$d(r) = 0, \quad r > \frac{p}{m}.$$
- Cases B.I and B.II.a tell us that for $\frac{p}{m} \leq \frac{n}{2n-1}$,
$$d(r) \leq n\left(1-\frac{mr}{p}\right), \quad 0 \leq r \leq \frac{p}{m}.$$
This upper bound improves for higher values of $\frac{p}{m}$. Hence, we must choose the maximum possible value of $\frac{p}{m}$ in the permissible range which is $\frac{n}{2n-1}$.
- From Cases B.I and B.II.b, we can see that for $\frac{p}{m} \geq \frac{n}{2n-1}$,
$$d(r) \leq \begin{cases} n\left(1-\frac{(n-1)mr}{nq}\right), & 0 \leq r \leq \frac{q}{m} \\ \frac{p}{p-q}\left(1-\frac{mr}{p}\right), & \frac{q}{m} \leq r \leq \frac{p}{m} \end{cases}.$$
We can see that the best tradeoff occurs for the choice $\frac{p}{m} = \frac{n}{2n-1}$ and this tradeoff coincides with the best tradeoff obtained from cases B.I and B.II.a.
- We always have the choice of avoiding relay cooperation, and thus of achieving $d(r) = (1-r)$ for any value of $r$.

With all the above observations, we get the statement of Theorem 2.

## Appendix III
## Solution of the Optimization problem for the NSDF Protocol

Let
$$\begin{aligned} d_a(r) &= (n-1)\left(1-\frac{mr}{p}\right)^+ + (1-r)^+ \\ &= \begin{cases} n\left(1-\frac{(n-1)m+p}{np}r\right), & 0 \leq r \leq \frac{p}{m} \\ 1-r, & \frac{p}{m} \leq r \leq 1 \end{cases} \end{aligned}$$
and
$$\begin{aligned} d_b(r) &= \min_{2\leq k\leq n}\left((n-k)\left(1-\frac{mr}{p}\right)^+ + d_k(r)\right), \\ &\qquad 0 \leq r \leq \frac{p}{m}, \end{aligned} \tag{42}$$
where $d_k(r)$ is given in (31). Therefore,
$$d(r) = \min\{\, d_a(r), d_b(r) \,\}, \quad 0 \leq r \leq \frac{p}{m}. \tag{43}$$
Also, from (36), we know that the DMT of the fixed-NSDF protocol is $(1-r)^+$ for $r > \frac{p}{m}$. Since the above expression for the DMT also evaluates to $(1-r)^+$ for $r > \frac{p}{m}$, for the sake of simplicity, we can assume that the expression in (43) is the actual DMT for all $r$.

Substituting the value of $d_k(r)$ in (42), we can see that the minimum occurs when $k = n$. Therefore, for the case $p \geq q$, we have
$$d_b(r) = d_n(r) = \begin{cases} n\left(1-\frac{(n-1)m}{nq}r\right), & 0 \leq r \leq \frac{q}{m} \\ \frac{m}{p}(1-r), & \frac{q}{m} \leq r \leq 1 \end{cases}.$$

Depending on the choice of the ratio $\kappa = \frac{p}{q}$ and the multiplexing gain $r$, either $d_a(r)$ or $d_b(r)$ will determine the actual DMT $d(r)$. It can be shown that there is a critical value of $\kappa$, which we shall denote by $\kappa_n$, below which $d(r) = d_a(r)$, i.e.,
$$d(r) = d_a(r), \quad 1 \leq \kappa \leq \kappa_n.$$
In order to compute $\kappa_n$, we compare the curves corresponding to $d_a(r)$ and $d_b(r)$. We can see that at $\kappa = \kappa_n$,
$$\begin{aligned} \frac{(n-1)m+p}{np} &= \frac{(n-1)m}{nq} \\ pnm - pm &= qnm - qm + pq \end{aligned}$$
By substituting $\kappa_n = \frac{p}{q}$ in the above equation, we get
$$(n-1)\kappa_n^2 - \kappa_n - (n-1) = 0.$$
Hence,
$$\kappa_n = \frac{1+\sqrt{1+4(n-1)^2}}{2(n-1)}.$$

Therefore, when $\kappa < \kappa_n$, the DMT is as given in (27). When $\kappa \geq \kappa_n$, both $d_a(r)$ and $d_b(r)$ determine the DMT for different ranges of $r$. The tradeoff is as given in (28). We can check that the point of intersection of $d_a(r)$ and $d_b(r)$, when $\kappa \geq \kappa_n$, is given by
$$r = \frac{np-m}{(n-2)m+p} \tag{44}$$

and

$$d(r) \;=\; \frac{m}{p}\left(1-\frac{np-m}{(n-2)m+p}\right). \qquad (45)$$

Hence, we have computed the DMT for the fixed-NSDF protocol.

Now, to get the best possible DMT in case of the variable-NSDF protocol, we vary $\kappa$ with the multiplexing gain $r$, i.e., we choose the value of $\kappa$ which maximizes the diversity at any given multiplexing gain. By comparing the DMTs in (27) and (28), it is clear that we must choose $\kappa = \kappa_n$ for $r \leq \frac{1}{\kappa_n+1}$. For $r \geq \frac{1}{\kappa_n+1}$, we need to track the point of intersection of $d_a(r)$ and $d_b(r)$. This point would correspond to maximum diversity at a certain $r$. By substituting $\kappa = \frac{p}{m}$ in (44) and (45), and by eliminating $\kappa$ from both the equations, we get

$$d(r) \;=\; \frac{(n-r)(1-r)}{(n-2)r+1}\;, \quad \frac{1}{\kappa_n+1} \;\leq\; r \;\leq 1.$$

With all the results mentioned above, we get the statement of Theorem 5.

## Acknowledgements

Thanks are due to Pankaj Bhambhani for helping out with the optimization in Theorem 2.